\documentclass[aps,prd,twocolumn,amsmath,amssymb,nofootinbib]{revtex4-1}
\usepackage{graphicx}
\usepackage[usenames,dvipsnames]{color}
\usepackage{times}
\usepackage{inputenc}
\usepackage{bm}
\usepackage{ulem}
\usepackage{multirow}
\usepackage{float}
\usepackage{url}
\usepackage{natbib}
\usepackage{subfigure}
\usepackage[colorlinks=true,citecolor=BrickRed,urlcolor=BrickRed,linkcolor=BrickRed]{hyperref}
\begin{document}
\preprint{}

\title{Non-radial oscillations in anisotropic dark energy stars}
\author{O. P. Jyothilakshmi}
\email{op\_jyothilakshmi@cb.students.amrita.edu}
\author{Lakshmi J. Naik}
\email{jn\_lakshmi@cb.students.amrita.edu}
\author{V. Sreekanth}
\email{v\_sreekanth@cb.amrita.edu}
\affiliation{Department of Physics, Amrita School of Physical Sciences, Amrita Vishwa Vidyapeetham, Coimbatore, India}

\date{\today}
\begin{abstract} 
We study the non-radial $f$-mode oscillations of both isotropic and anisotropic dark energy stars by using the modified Chaplygin prescription of dark energy to model the stellar matter. The anisotropic pressure in the system is modeled with Bowers-Liang prescription. By solving the stellar structure equations in presence of anisotropy, we study the global properties of the dark energy star and compare the mass-radius profiles with data from GW events and milli-second pulsars. We proceed to determine the prominent non-radial $l=2$ $f$-mode frequencies of the anisotropic dark energy star by employing the Cowling approximation and analyse and quantify the spectra by varying the anisotropic parameter. We report that $f$-mode spectra of dark energy star have distinctly different behaviour compared to neutron star and quark star, and this may possibly help in its future identification. Further, the tidal deformability factors of the anisotropic dark energy stars have also been analyzed.

\end{abstract}
\maketitle

\section{Introduction}
\label{sec:intro}
Dark energy is the fluid component that governs the accelerated expansion of the universe \cite{SupernovaSearchTeam:1998fmf,SupernovaCosmologyProject:1998vns,Freedman:2003ys}. It is found from various observations that dark energy fills up almost $70\%$ of the universe \cite{WMAP:2008ydk}. 
The $\Lambda$CDM model (also known as the concordance model) \cite{Einstein:1917ce} is one of the most accepted cosmological models describing dark energy. 
However, the $\Lambda$CDM model suffers from problems like the cosmological constant problem \cite{Weinberg:1988cp,Zeldovich:1967gd},  
and the Hubble tension \cite{Mortsell:2018mfj,Verde:2013wza,Bolejko:2017fos}. 
Therefore, many alternative models have been proposed in the last few decades. A detailed review on various dynamical models of dark energy such as cosmological constant, quintessence, K-essence, tachyon field, phantom field, dilatonic field, Chaplygin gas model, etc. are given in Ref. \cite{Copeland:2006wr}. 
\par 
The fact that interior composition of a compact star is unknown to this date has motivated a lot of researchers to construct different models of compact stars. There are several stellar models that include dark energy such as 
false vacuum bubbles~\cite{Coleman:1980aw}, 
non-singular black holes \cite{Dymnikova:1992ux}, gravastars \cite{Mazur:2001fv}, and dark energy stars \cite{Chapline:2004jfp}.
Among these models, we are interested in the dark energy star, which was first proposed by Chapline~\cite{Chapline:2004jfp}. The author gave an alternative explanation to astrophysically observed black holes suggesting that these compact objects may be dark energy stars. The author also proposed that matter is converted to dark energy when it falls through the event horizon. 

Using the general relativistic prescriptions, the structure of compact stars can be obtained by describing the interior composition with an equation of state (EoS). Thus obtained global properties of compact objects are compared with the astrophysical observations to constrain the EoS.
Recently detected gravitational waves (GWs) from binary neutron star mergers introduced a new way of restricting the dense matter EoS within neutron stars \cite{LIGOScientific:2018cki}. The constraints of EoS also include the maximum mass limit set by various observational data from pulsars \cite{NANOGrav:2019jur,Antoniadis2013}.
GW asteroseismology is the study of the interior of compact objects using GW observations. The stellar oscillations can produce GWs, this has drawn a great interest recently in the field of GW asteroseismology.
The analysis of stellar oscillations is used to understand the microscopic and macroscopic properties of compact objects~\cite{Lindblom:2001hd,Jha10,PhysRevD.83.024014,Doneva:2012rd,Flores:2013yqa,Jyothilakshmi:2022hys,Tran:2022dva}. The non-radial modes of stellar oscillations are an important class among these and the general relativistic treatment of the same can be found in Ref.~\cite{Thorne,Detweiler:1985zz}. 

These stellar oscillations can be classified into various quasi-normal modes based on the force that restores the system back to equilibrium. Some examples include fundamental $f$-mode, pressure $p$-mode, gravity $g$-mode, rotational $r$-mode, and space-time $w$-mode~\cite{Kokkotas:1999bd}. Among these modes, $f$-modes are considered to be a promising candidate for GW emissions, which has frequencies in the sensitivity range of upcoming GW detectors~\cite{Andersson:2002ch}. 

Another important constraint imposed on EoS is by the tidal distortions caused from binary neutron star inspiral \cite{Hinderer2008,Hind2010}. 
The tidal deformability parameter $\Lambda$ describes the gravitational wave signal emitted during a binary neutron star inspiral~\cite{Hinderer2008,Hind2010}. It determines a star's quadrupole deformation caused by its companion star's tidal field. 

This is in addition to the constraints previously provided by the electromagnetic studies of neutron stars, which include their masses, radii, spin, and gravitational redshift.
The tidal properties of neutron stars can be measured from the GW signals \cite{Annala:2017llu}.
\par 
In most of the astrophysical studies, the fluid matter within dense compact stars is assumed to be locally isotropic.
However, the high density and strong gravity suggest that the interior of compact stars could be anisotropic, meaning the pressure in radial and tangential directions are different. Strong magnetic field, superfluid cores, phase transition, pion condensation etc. may give rise to pressure anisotropy. 
Ruderman in 1972 was the first to propose the idea of anisotropy in compact stars \cite{Ruderman:1972aj} and later, the relativistic stellar structure equations for anisotropic stars were obtained by Bowers and Liang~\cite{Bowers:1974tgi}. 

Apart from this, various other anisotropy models are used in the study of compact stars \cite{Horvat:2010xf,Herrera:2013fja,Raposo:2018rjn}.
See Ref. \cite{KUMAR2022101662} for a recent review on anisotropic compact stars. 

Non-radial oscillations were studied in anisotropic neutron stars \cite{Doneva:2012rd} and recently, an extension of this work was done for hadronic and quark stars \cite{Curi:2022nnt}. These studies show that presence of anisotropy has significant effect on non-radial oscillations. Further, the impact of anisotropy on the tidal deformability was also studied~\cite{Biswas:2019gkw}. It will be interesting to study these effects in dark energy stars. 
There has been a considerable interest to study the structure and properties of compact objects with dark energy. 
In Ref. \cite{Lobo:2005uf}, the author introduced a model of dark energy star 
and studied its stability. Anisotropic dark energy stars were considered in Ref. \cite{Chan:2008ui}. 
Another model of dark energy star which includes neutron gas was constructed and the stability of the model was also analysed~\cite{Ghezzi:2009ct}. 
A mixed dark energy star model made up of baryonic matter and phantom scalar field has also been considered~\cite{Yazadjiev:2011sm}. Further, a stable singularity-free anisotropic dark energy star was constructed in Ref. \cite{Rahaman:2011hd}.

A time-dependent solution of the Einstein field equations describing the collapse of a spherical system resulting in a dark energy core was obtained in Ref.~\cite{Beltracchi:2018ait}. The stability of a dark energy star with a phantom field was also investigated \cite{Sakti:2021mvd}.  Recently, a study on dark energy star within Newtonian treatment was performed with the generalised Chaplygin EoS by including the effects of anisotropy 
\cite{Abellan:2023tft}. 
Global properties of slowly rotating isotropic dark energy stars with extended Chaplygin EoS were analyzed in Ref.~\cite{Panotopoulos:2021dtu}.
Radial oscillations and tidal Love numbers using generalised Chaplygin EoS prescription was performed
for the isotropic \cite{Panotopoulos:2020kgl} and anisotropic \cite{Pretel:2023nhf} dark energy stars. Further, it was shown that a neutron star with dark energy core is dynamically stable under small radial pulsations and is consistent with observational data \cite{Pretel:2024tjw}. Dark energy stars within modified theories of gravity have also drawn interest recently \cite{Das:2023bff,BagheriTudeshki:2023dbm}. 
In this work, we intend to study the prominent non-radial $f$-mode oscillations in both isotropic and anisotropic dark energy stars constructed using the modified Chaplygin EoS. The anisotropy is introduced via the Bowers-Liang model \cite{Bowers:1974tgi}. We also calculate tidal deformability for such a system.

The paper is organised as follows. In Section~\ref{Sec:EoS}, we discuss the modified Chaplygin EoS describing dark energy used for the analysis. We present the stellar structure equations and non-radial oscillations for anisotropic dark energy star in Section~\ref{Sec:non-radial}. The results of our analysis are  shown in Section~\ref{sec:results} and finally we draw the conclusions of the study and summarise in Section~\ref{sec:summ}. 

\textit{Notations and conventions}: Throughout this paper, we set  Newton's universal gravitational constant $G$ and velocity of light in free space $c$ as $G=c=1$ and follow the metric convention $g_{\mu\nu}=\textrm{diag}(-1,1,1,1)$.

\section{Chaplygin Gas Model of Dark Energy}\label{Sec:EoS}

The Chaplygin fluid model of dark energy was obtained by Kamenshchik et al. in 2001~\cite{Kamenshchik:2001cp}, where the universe is assumed to be filled with the so called \textit{Chaplygin gas} obeying the equation of state (EoS) 
\begin{equation}\label{Eq:Chap-1}
    p = -\hat{B}/\rho^\omega,
\end{equation}
where $p$ is the pressure, $\rho$ is the energy density and $\hat{B}$ is a positive constant with units of energy density and $\omega$ can take values in the range $0<\omega\leq 1$. 

This model provides a possible solution for the unification of dark matter and dark energy \cite{Bilic:2001cg,Bento:2002ps,Gorini:2002kf,Xu_2012}. A more generalised version of the Chaplygin EoS was introduced in Ref. \cite{Zhang:2004gc}. Furthermore, a modified Chaplygin EoS was later constructed by including the effects of viscosity \cite{Pourhassan:2013sw,Saadat:2013ava} 
\begin{equation}\label{Eq:Chap-2}
p=\hat{A}\rho-\frac{\hat{B}}{\rho^{\omega}},
\end{equation}%
where $\hat{A}$ is a dimensionless positive constant. The first term represents a barotropic and the second term corresponds to the Chaplygin gas. 
By setting the value of $\omega=1$, one obtains the modified Chaplygin EoS in the form~\cite{Kahya:2015dpa}
\begin{align}
p=A^2 \rho - \frac{B^2}{\rho}. \label{EOS}
\end{align} 
Since the pressure vanishes at the surface of the star, the energy density towards the surface takes the value of $\rho_s=B/A$. This EoS is used in various studies on dark energy stars ~\cite{Panotopoulos:2020kgl,Panotopoulos:2021dtu,Pretel:2023nhf,Abellan:2023tft,Das:2023bff,BagheriTudeshki:2023dbm,Pretel:2024tjw} and 
following Refs. \cite{Panotopoulos:2021dtu,Panotopoulos:2020kgl,Pretel:2023nhf}, we take the values of the constants as $A=\sqrt{0.4}$ and $B=0.23\times 10^{-3}$/km$^2$.

Next, we proceed to study the non-radial $f$-mode oscillations of anisotropic dark energy star.

\section{Anisotropy and Non-radial oscillation}\label{Sec:non-radial}

 In this section, we describe the stellar structure equations, non-radial oscillations and tidal deformability for an anisotropic stellar system.
 The stress-energy tensor $T^{\mu\nu}$ for an anisotropic fluid is given by 
\begin{align}
\label{Tmunu}
        T^{\mu\nu}&=\rho u^\mu u^\nu + qg^{\mu\nu} +\sigma k^\mu k^\nu;
\end{align}
where $\sigma=p-q$ is the anisotropy pressure, $p$ is the radial pressure, $q$ is the tangential pressure, $\rho$ is the energy density, $u^\mu$ is the four-velocity of the fluid, and $k_\nu$ is the radial vector. The four-vectors satisfy the conditions: $u_\mu u^\mu =-1$, $k_{\mu}k^{\mu}=1$ and $u^\mu k_{\mu}=0$ . We use the Bowers-Liang model for anisotropy~\cite{Bowers:1974tgi}, 
\begin{align}\label{Eq:Sigma}
    \sigma=-\lambda_{BL}\frac{ r^3}{3}\left(\rho(r)+3p(r)\right)\left(\frac{\rho(r) + p(r)}{r-2m(r)}\right);
\end{align}
where free parameter $\lambda_{BL}$ denotes the measure of anisotropy. Following Ref.~\cite{Biswas:2019gkw}, we consider $-2<\lambda_{BL}<2$ in this work. 
The line element for a spherically symmetric static star in Schwarzschild coordinates, ($t,\, r,\, \theta,\, \phi$) is given by
\begin{align}\label{metric}
   ds^{2} &= -e^{2\Phi} dt^{2}+e^{2 \lambda} dr^{2}+r^{2}\left(d\theta ^{2}+\sin ^{2}\theta d\phi ^{2}\right).
\end{align}
Here, $\Phi(r)$ and $\lambda(r)$ are the metric functions.
Solving the Einstein field equations for the static spherically symmetric metric for an anisotropic fluid results in the 
stellar structure equations
\cite{Bowers:1974tgi} 
\begin{subequations}
  \begin{align}\label{TOV-1}
    \frac{dp}{dr} &=-\frac{(\rho+p)(m+4\pi r^3 p)}{r^2(1-2m/r)}-\frac{2\sigma}{r},\\
    \label{TOV-2}
    \frac{dm}{dr} &= 4\pi r^2 \rho.
  \end{align}   
\end{subequations} 
These equations reduce to the well known Tolman-Oppenheimer-Volkoff (TOV) equations \cite{Tolman:1939jz,Oppenheimer:1939ne} of the isotropic case, if the value of anisotropic factor is set to zero $(\sigma=0)$.
The above shown coupled differential equations are integrated from the center to the surface of the star for a given EoS describing the stellar matter. The initial conditions towards the center of the star are pressure $p_c = p(0)=p(\rho_c)$ and mass $m_c =m(0)= 0$. The pressure $p$ tends to zero as it approaches the surface of the star ($r=R$) and the mass of the star is then obtained as $M=m(R)$. For a given EoS, we can obtain stellar configurations with different masses and radii by varying the value of central density $\rho_c$.

The solutions obtained by solving Eqs. (\ref{TOV-1}) and (\ref{TOV-2}) are used to study the non-radial oscillations of anisotropic dark energy stars. Here, we employ the well known Cowling approximation which assumes an unperturbed space-time while studying the fluid pertubations \cite{Cowling:1941nqk}. It was found that the results obtained using this approximation differ from the general relativistic solutions only by less than $20\%$ \cite{Sotani:2020mwc}. The oscillation equations in the Cowling approximation are obtained by considering the perturbation of energy-momentum tensor conservation ~\cite{Flores:2013yqa,Doneva:2012rd,PhysRevD.83.024014}.

The system of differential equations in terms of the fluid perturbations  $V(r)$ and $W(r)$ representing the anisotropic non-radial oscillations are given by \cite{Doneva:2012rd}
\begin{subequations}
\begin{eqnarray}
\frac{d W}{dr}&=& \frac{d\rho}{dp}\Big[\omega^2 \frac{\rho + p-\sigma}{\rho + p} \left(1 - \frac{\partial\sigma}{\partial
p}\right)^{-1}e^{\lambda - 2\Phi} r^2 V \label{W-Cowling}   \nonumber \\
&& + \frac{d\Phi}{dr} W\Big] - l(l+1)e^{\lambda} V    
\\ \nonumber
&&+\frac{\sigma}{\rho + p}\left[\frac{2}{r}\left(1+ \frac{d\rho}{dp}\right)W + l(l+1)e^{\lambda}V \right],\\
\frac{dV}{dr} &=& 2V \frac{d\Phi}{dr} - \left(1 -\frac{\partial \sigma}{\partial p} \right) \frac{\rho + p}{\rho + p-\sigma}
\frac{e^{\lambda}}{r^2} W  \label{V-Cowling} \\
  &+& \Big[\frac{\sigma^{\,\prime}}{\rho +
p -\sigma} + \left(\frac{d\rho}{dp}+ 1\right)\frac{\sigma}{\rho + p-\sigma} \left(\frac{d\Phi}{dr} + \frac{2}{r}\right)
\nonumber \\ \nonumber
&-& \frac{2}{r}\frac{\partial\sigma}{\partial p} - \left(1 - \frac{\partial\sigma}{\partial
p}\right)^{-1}\left(\frac{\partial^{2}\sigma}{\partial p^2} p^\prime + \frac{\partial^2\sigma }{\partial p\partial
\mu}\mu^\prime\right)\Big] V .\nonumber
\end{eqnarray}
\end{subequations}
Here, $\mu = 2m/r=1-e^{-2\lambda}$. 
These differential equations are solved from the center to the surface of the star, which obey the following conditions as $r \rightarrow 0$:
\begin{eqnarray}\label{BC-I}
W= A r^{l+1}, \;\;\; V= -\frac{A}{l} r^l; 
\end{eqnarray}
where $A$ is an arbitrary constant.
The boundary condition towards the surface of the star is \cite{Doneva:2012rd}
\begin{eqnarray}\label{BC-II}
\omega^2 \left(1 - \frac{ \sigma}{\rho + p}\right)\left( 1-\frac{\partial\sigma}{\partial p}\right)^{-1}e^{-2\Phi}V \nonumber\\+ \left(\frac{d\Phi}{dr}  + \frac{2}{r} \frac{\sigma}{\rho + p}\right)e^{-\lambda}\frac{W}{r^2}=0.
\end{eqnarray}
The coupled differential equations Eqs. (\ref{W-Cowling}) and (\ref{V-Cowling}) are integrated using the boundary conditions given by Eqs. (\ref{BC-I}) and (\ref{BC-II}). 
Initially, we assume a value for $\omega ^2$ and after each integration, the value of $\omega^2$ is modified until the surface boundary condition is satisfied.

\begin{figure}[t]
\centering
\includegraphics[width=0.5\textwidth]{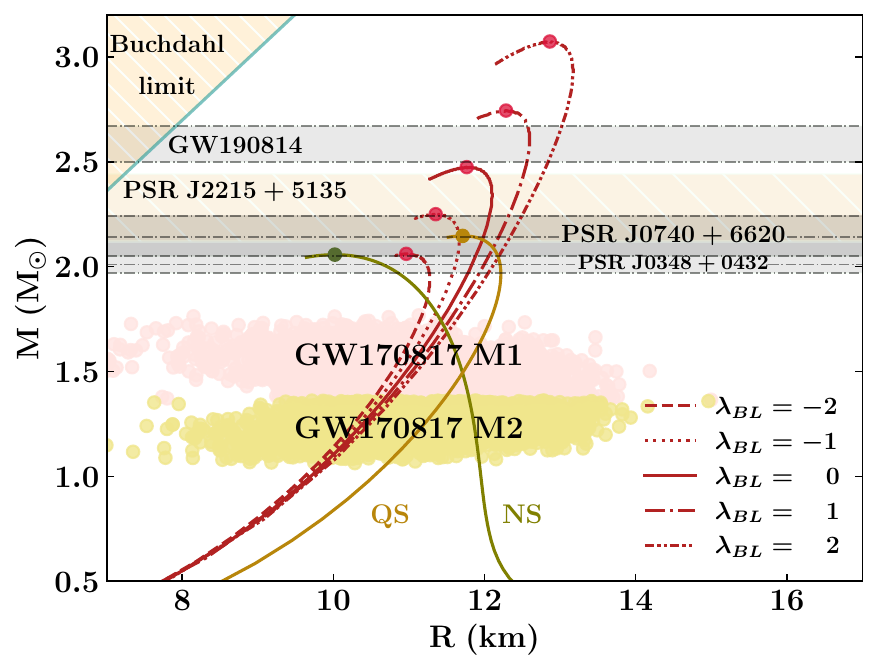}
\caption{Mass-radius curve of static anisotropic dark energy stars with $\lambda_{BL}$ ranging from $-2$ to $+2$. $\lambda_{BL}=0$ corresponds to the isotropic case. The mass-radius curves of isotropic neutron star (NS), with Sly4 EoS and quark star (QS), with modified Bag model EoS are also shown.  
The horizontal bands are observational constraints from GW 190814 events \cite{Abbott2020} and various pulsars (PSR J2215+5135 ($M = 2.27_{-0.15}^{+0.17} M_\odot$) \cite{Linares:2018ppq}, PSR J0348+0432 ($M = 2.01_{-0.04}^{+0.04} M_\odot$) \cite{Antoniadis2013} and PSR J0740+6620 ($2.14_{-0.09}^{+0.10} M_\odot$ ($68.3\% $ credible) ) \cite{NANOGrav:2019jur}). We indicate the observational measurements from GW 170814 event \cite{LIGOScientific:2018cki}. The Buchdahl limit ($2M/R \leq 0.88$) \cite{Buchdahl}  is also shown here.}
\label{fig:M-R}
\end{figure}
To obtain the tidal perturbations in presence of anisotropy, we follow the prescription developed in Ref. \cite{Biswas:2019gkw} based on  Refs.~\cite{Hinderer2008,Hind2010}.

The tidal deformability $\Lambda$ of a static, spherically symmetric star placed in a static external quadrupolar tidal field $\mathcal{E}_{ij}$ 
is defined in  terms of induced quadrupole moment $Q_{ij}$ as $\Lambda = -Q_{ij}/\mathcal{E}_{ij}$ and can be expressed in terms of stellar radius $R$, mass $M$ and tidal Love number $k_2$ as $\Lambda = 2k_2R^5/3$~\cite{Hinderer2008,Hind2010}. 
In order to study the tidal perturbations, we use the set of 
coupled differential equations in terms of the fluid perturbation function $H$ given by \cite{Biswas:2019gkw}
\begin{eqnarray}
\frac{dH}{dr}&=& \beta,\label{Tidal-1}\\
  \frac{d\beta}{dr}&=& \frac{2rH}{r-2m}\left\{-2\pi
  \left[4\rho+8p + \frac{\rho+p}{\mathcal{A}c_{s}^2}(1+c_s^2)\right]\phantom{\frac{3}{r^2}} \right. \nonumber\\
&+& \left. \frac{3}{r^2}+\frac{2r}{r-2m}
  \left(\frac{m}{r^2}+4\pi r p\right)^2\right\}\nonumber\\
&+&\frac{2\beta}{r-2m}\left\{-1+\frac{m}{r}+2\pi r^2
  (\rho-p)\right\}.\label{Tidal-2}
\end{eqnarray}
Here, $\mathcal{A}=dq/dp$ and $c_s^2=dp/d\rho$ is the speed of sound squared. By setting $\mathcal{A}=1$, the above equations reduce to the isotropic case ~\cite{Hinderer2008,Hind2010}. The integration is performed outwards starting from the stellar center ($r\to0$) with the values $H(r) = a_0r^2$ and $\beta(r) = 2a_0r$, where $a_0$ is an arbitrary constant. 
The interior and exterior solutions are matched at the surface of the star and the solution obtained can be used to calculate the $l=2$ tidal Love number $k_2$ ~\cite{Hinderer2008,Hind2010}
\begin{eqnarray}\label{eq:k2}
k_2 &=& \frac{8C^5}{5}(1-2C)^2[2+2C(y-1)-y]\nonumber\\
      && \times\bigg\{2C[6-3y+3C(5y-8)]\\
      && +4C^3[13-11y+C(3y-2)+2C^2(1+y)]\nonumber\\
      &&+3(1-2C)^2[2-y+2C(y-1)] \ln(1-2C)\bigg\}^{-1}; \nonumber
\end{eqnarray}
where $y=R\beta(R)/H(R)$ and compactness of the star $C=M/R$, which gives a measure of the strength of star's gravity.

\section{Results and discussions}
\label{sec:results}

Now, we proceed to obtain the stellar configurations, non-radial $f$-mode oscillations and tidal deformability factors of both isotropic and anisotropic dark energy stars that obey the modified Chaplygin EoS discussed in Sec.~\ref{Sec:EoS}. 
We begin our analysis by numerically solving the static stellar structure equations: Eqs. \eqref{TOV-1} and \eqref{TOV-2} from the center to the surface of the star with the appropriate boundary conditions given in Sec.~\ref{Sec:non-radial}.  The $f$-mode frequencies are calculated by 
numerically solving the coupled differential equations: Eqs. \eqref{W-Cowling} and \eqref{V-Cowling} with the boundary conditions given by Eqs. \eqref{BC-I} and \eqref{BC-II}. 
Here, we use Ridder's method to obtain the eigen or $f$-mode frequency with sufficient precision. 
Finally, we obtain the tidal deformability $\Lambda$ by solving the second-order differential equation Eq.~\eqref{Tidal-2} from the center to the surface of the star. 
We also compare our results with that obtained for isotropic neutron and quark stars. For neutron star (NS) matter, we consider the Sly4 EoS \cite{Douchin:2001sv}
; whereas for the (strange) quark star (QS), the modified Bag model \cite{Fraga:2001id} is used. 

In Fig.~\ref{fig:M-R}, we plot the mass-radius profiles of the anisotropic dark energy stars for different values of anisotropic strength. 
Following Ref.~\cite{Biswas:2019gkw}, we consider the range: $-2<\lambda_{BL}<2$ in this work. Here, $\lambda_{BL}=0$ corresponds to the isotropic stellar profile. 
As $\lambda_{BL}$ varies positively (negatively) from the isotropic value, both mass and radius of the star increase (decrease). This trend is in agreement with Ref. \cite{Pretel:2023nhf}, although a different model of anisotropy was employed in their analysis of dark energy stars.
The maximum mass values corresponding to $\lambda_{BL} = -2,\,-1,\,0,\,1,$ and $2 $ are obtained as $2.06,\,2.25,\,2.47,\,2.74,$ and $3.07 M_\odot$ respectively; the corresponding central density values are $(2.24,\,1.85,\,1.69,\,1.43,$ and $1.22) \times 10^{15}$ g cm$^{-3}$. 
Thus, we see that anisotropy has significant impact on the maximum mass and radius of the stellar profiles. 
We note that the effect of anisotropy on the mass and radius values are more prominent at higher central densities. 
Further, one can see that neutron star profiles are different: results in lower value of maximum mass and corresponding radius and higher radius for lower central densities. 
However, we see that the isotropic quark star profile is qualitatively similar to that of the dark energy star; the maximum mass and radius values are lesser than that of the dark star. The maximum mass (corresponding radius) of NS and QS are $2.06 M_\odot$ ($10.02$ km) and $2.15\, M_\odot$ ($11.71$ km) respectively. 
\par 
We now compare the dark energy star results shown in Fig. \ref{fig:M-R} with some of the observational measurements given by GW events and milli-second pulsars. We have shown the observational data from GW 170817 \cite{LIGOScientific:2018cki} and GW 190814 \cite{Abbott2020} events. We have also indicated the observational limits from PSR J2215+5135 ($M = 2.27_{-0.15}^{+0.17} M_\odot$) \cite{Linares:2018ppq}, PSR J0348+0432 ($M = 2.01_{-0.04}^{+0.04} M_\odot$) \cite{Antoniadis2013} and PSR J0740+6620 ($2.14_{-0.09}^{+0.10} M_\odot$ ($68.3\% $ credible)) 
 \cite{NANOGrav:2019jur}. We find that the maximum masses of anisotropic dark energy stars with $\lambda_{BL}$ values $-1$ and $-2$ are consistent with the PSR J0740+6620, PSR J0348+0432 and PSR J2215+5135. We also note that the mass-radius values of anisotropic dark energy stars fall within the limits imposed by GW 170817; whereas, the values of mass and radius obey the limit imposed by GW 190814 only for $\lambda_{BL}=-1$ and $-2$. 
Furthermore, we have shown the Buchdahl limit \cite{Buchdahl} in Fig. \ref{fig:M-R} and it can be seen that the anisotropic dark energy star profiles considered in this work fall well within this limit.


\begin{figure*}
  \centering
  \subfigure
  []{\includegraphics[height=22cm]{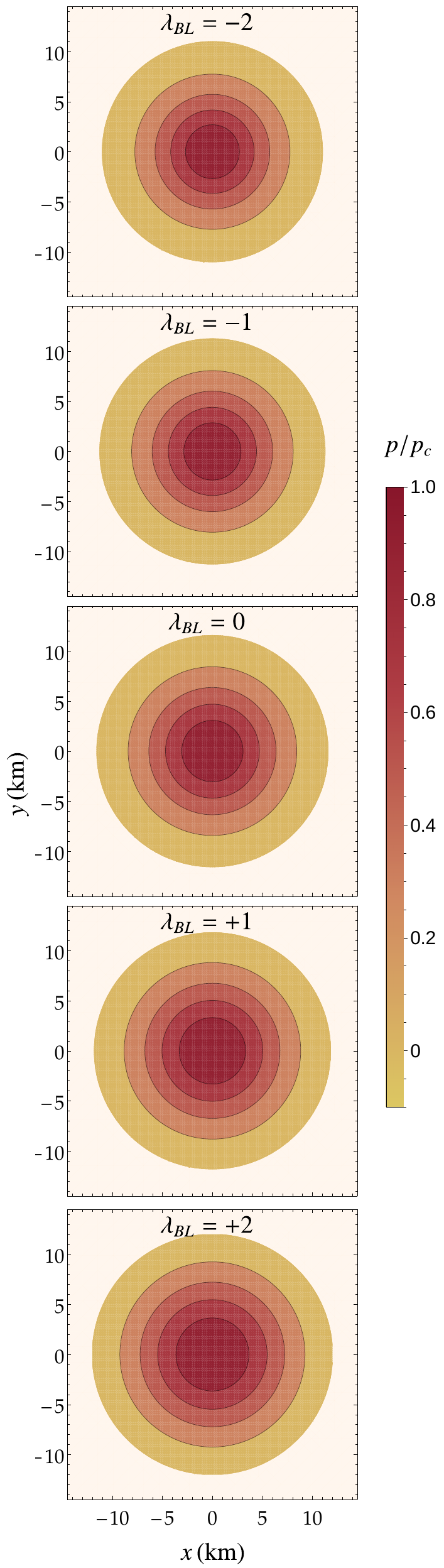}\label{fig:radial}}\quad\qquad\quad
  \subfigure[]{\includegraphics[height=22cm]{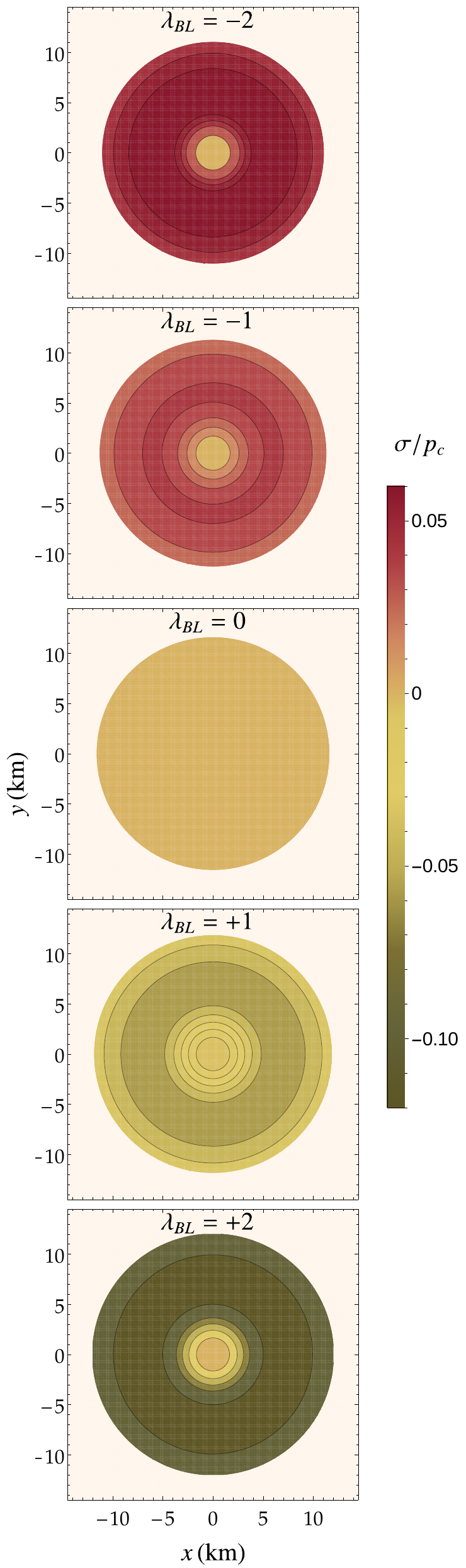}\label{fig:aniso} 
  } 
  \caption{The scaled (a) radial pressure $p/p_c$ and (b) anisotropic pressure $\sigma/p_c$ as a function of radial co-ordinate $r=\sqrt{x^2+y^2}$ in $z=0$ plane of the anisotropic dark energy star with central pressure $p_c=6.75\times 10^{35}$ g cm$^{-1}$s$^{-2}$. 
  }
  \label{fig:contour}
  \end{figure*}
\begin{figure*}
  \centering
  \subfigure
  []{\includegraphics[width=0.47\linewidth]{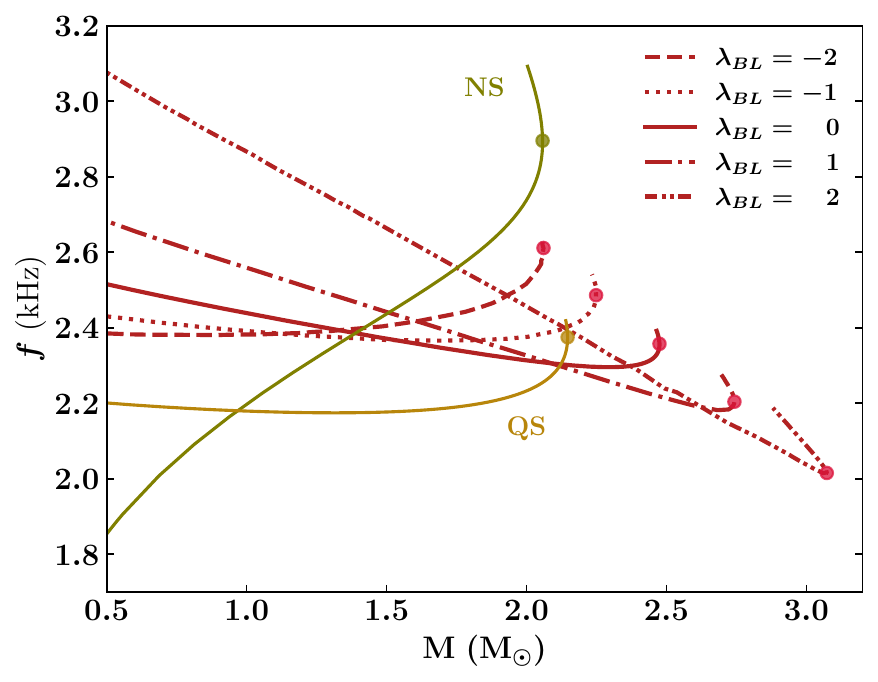}\label{fig:f-M}}\quad
  \subfigure[]{\includegraphics[width=0.47\linewidth]{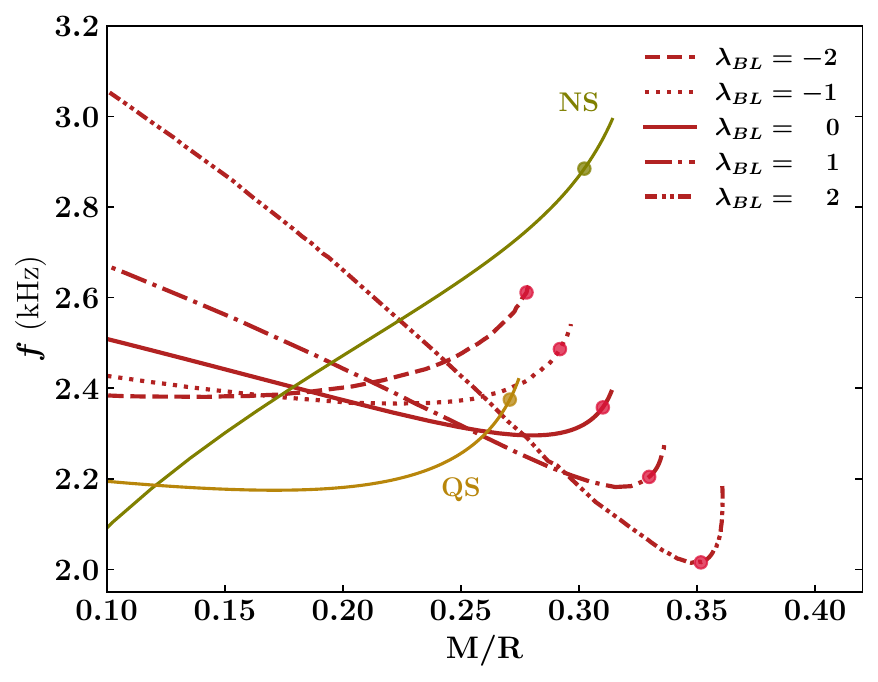}\label{fig:f-C} 
  } 
  \caption{$f$-mode frequency as a function of (a) mass and (b) compactness, $C=M/R$ of the anisotropic dark energy star for $l=2$ mode. The results for neutron star (NS) and quark star (QS) are also depicted.}
  \label{fig:f-M-C}
  \end{figure*}
\begin{figure*}
\centering
\subfigure[]{\includegraphics[width=0.48\linewidth,height=7cm]{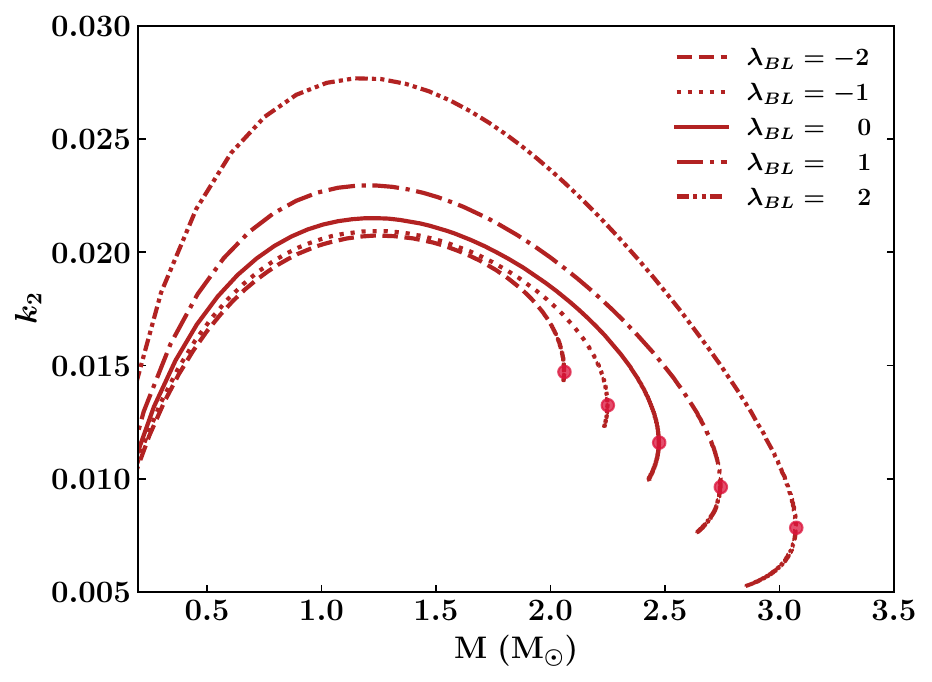}\label{fig:k2-M}}\quad
\subfigure
[]{\includegraphics[width=0.48\linewidth,height=7cm]{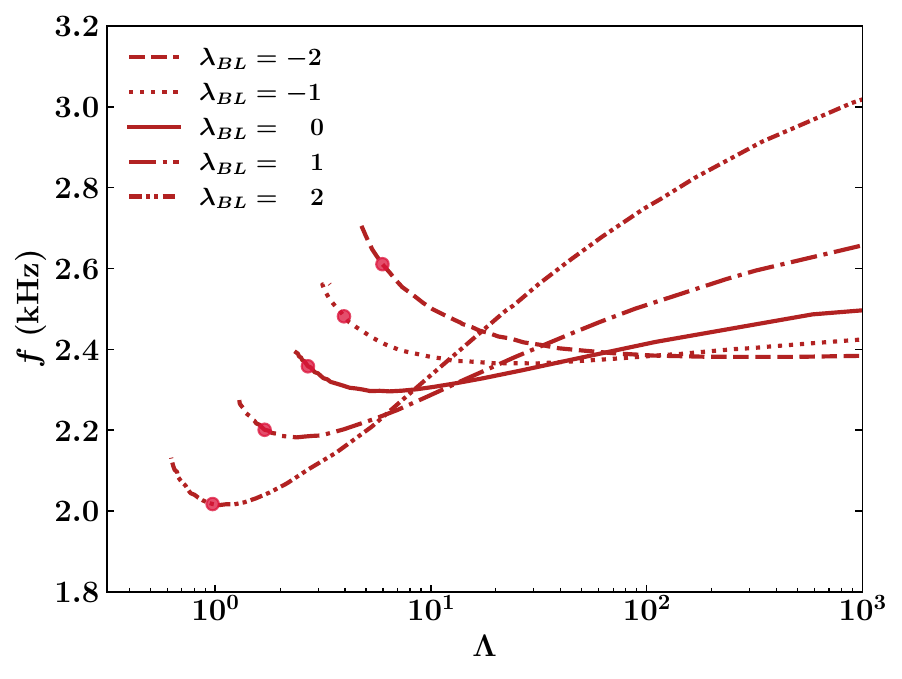}\label{fig:L-f}}
\caption{The (a) tidal Love number $k_2$ against mass of the anisotropic dark energy star and (b) $f$-mode frequency as a function of tidal deformability $\Lambda$.}
\label{fig:f1}
\end{figure*}
Next, we present the variation of scaled radial pressure $p/p_c$ and anisotropic pressure $\sigma/p_c$ of dark energy stars as a function of radial co-ordinate $r=\sqrt{x^2 + y^2}$ with $z=0$ for different values of $\lambda_{BL}$ in Fig.~\ref{fig:contour}. The stellar profiles are obtained by keeping a fixed value of central pressure $p_c=6.75\times10^{35}$ gcm$^{-1}$s$^{-2}$, with corresponding central density being $2.0\times 10^{15}$ g cm$^{-3}$. 
The radius of the star increases marginally with an increase in the value of $\lambda_{BL}$. For $\lambda_{BL}=-2,-1,0,1,2$, the radius values are obtained as $11.04$, $11.29$, $11.59$, $11.84$, and $12.04$ km respectively. 
 The radial pressure for any given value of $r$ is found to increase (decrease) with respect to the isotropic case $i.e.,\, \lambda_{BL} = 0$, with increase (decrease) in the value of $\lambda_{BL}$, as shown in Fig.~\ref{fig:radial}. The variation of radial pressure from the isotropic case is maximum towards the mid of the radii of the profiles. Now, coming to the anisotropic pressure profiles, we find that it changes drastically as the value of $\lambda_{BL}$ varies. 
From Fig.~\ref{fig:aniso}, it can be seen that the anisotropic pressure increases from $0$ ($\lambda_{BL} = 0$), reaches a maximum value and then slowly decreases with $r$ for positive values of $\lambda_{BL}$ and vice versa for negative values. 
Further, the anisotropic pressure profiles show an overall decrement with increase in the value of $\lambda_{BL}$ from $-2$ to $2$. 
It can be seen that, $\sigma/p_c$ becomes more and more negative with increasing values of $\lambda_{BL}$ in the positive direction; vice-versa with negative increment of $\lambda_{BL}$.  
We also note that, for positive values of $\lambda_{BL}$, the deviation from the isotropic case is higher compared to negative values of $\lambda_{BL}$. We note that qualitatively same trend is observed in the case of neutron stars with a different anisotropy prescription in Ref.~\cite{Doneva:2012rd}. 

For $f$-mode analysis, we employ the Cowling approximation discussed in Sec.~\ref{Sec:non-radial}. In Figs.~\ref{fig:f-M} and \ref{fig:f-C}, we plot the $l=2$ $f$-mode frequencies as a function of dark energy stellar mass and compactness $C=M/R$ respectively. 
We observe that, the $f$-mode oscillation frequencies of isotropic dark energy star ($\lambda_{BL} = 0$) lie within a smaller range compared to that of the isotropic NS considered. Also, the oscillation frequency corresponding to the maximum mass, $f_{max}$ of the dark energy star is observed at a lower value compared to that of the NS.
Now, considering the anisotropic dark energy star, 
the $f$-mode frequencies corresponding to different values of $\lambda_{BL}$ can be observed distinctly 
from the isotropic case. 
We find that the frequency range of $f$-modes expands for more positive values of $\lambda_{BL}$.
From Fig.~\ref{fig:f-M}, the $f$-modes of dark energy star lie in the range of $2.38-2.61$ kHz and $2.86-2.02$ kHz for $\lambda_{BL}=-2$ and $+2$ respectively, with the mass ranging from $1.00 M_\odot$ to the maximum mass of the star obtained for each case.
The deviation of the $f$-mode spectra from the isotropic case is more prominent for larger absolute values of $\lambda_{BL}$. This is because, the variation of anisotropic pressure of the star is observed to be large for higher values of $|\lambda_{BL}|$. 
Further, we note that the $f$-mode frequencies show only a small deviation from isotropic case for negative $\lambda_{BL}$ values. For example, the values of $f$ corresponding to $1.00 M_\odot$ are $2.39$, and $2.38$ kHz respectively for $\lambda_{BL}=-1,$ and $-2$; while the frequencies are obtained as $2.43,\,2.55$, and $2.86$ kHz for $\lambda_{BL}=0,\,1,$ and $2$ respectively with the same mass.
Also, the $f_{max}$ values are found to decrease with increase in the value of $\lambda_{BL}$. 
The values of $f_{max}$ for $\lambda_{BL}=-2,\,-1,\,0,\,1,$ and $2$ are $2.61,\,2.48,\,2.36,\,2.20,$ and $2.02$ kHz respectively. 
\par 
More importantly, we observe that the $f$-mode frequency spectrum of dark energy stars is distinctively different from that of the neutron stars (NS) and quark stars (QS). 
We note that, the $f$-mode frequency corresponding to positive $\lambda_{BL}$ start from larger values compared to the isotropic case and decrease, then approach $f_{max}$; while for negative $\lambda_{BL}$ values, the curves start from smaller frequency range, cross the isotropic curve and approach $f_{max}$. 
Moreover, we observe that the $f$-modes of dark energy star begin in the frequency range $\approx 2.4-3.1$ kHz, unlike the case of NS and QS for which the curve begins at lower frequencies. For NS, the values of $f$ increase with the stellar mass and $f_{max}$ occur at largest value in the $f$-mode frequency range obtained for the star.
On the other hand, the $f$-mode values of dark energy star for positive $\lambda_{BL}$ decrease with mass and $f_{max}$ values are obtained at lower frequency range.
Additionally, it can be noted that the $f$-mode curve of dark energy star appears similar to that of QS when $\lambda_{BL}$ becomes more negative; the curve remains almost constant when mass is increased, then rises suddenly when approaching the maximum mass. 

 The $f$-mode frequency as a function of compactness $C=M/R$ is given in Fig.~\ref{fig:f-C}. The values of compactness corresponding to maximum mass $C_{max}$ for dark energy stars are $0.28,\,0.29,\,0.31,\,0.33,$ and $0.35$ with $\lambda_{BL}=-2,\,-1,\,0,\,1,$ and $2$ respectively. The values of $C_{max}$ for NS and QS are $0.30$ and $0.27$ respectively. We note that, the compactness of isotropic dark energy stars is of the order of NS. However, a positive deviation from isotropy results in ultra-compactness, $C>1/3$~\cite{BRIyer_1985}.

Further, we study the tidal properties and plot the relation between $f$-mode and tidal deformability $\Lambda$ of anisotropic dark energy stars.  In Fig.~\ref{fig:k2-M}, we plot the tidal Love number $k_2$ as a function of the stellar mass. The value of $k_2$ increases until it reaches a maximum and then decreases as the mass is increased. A similar trend was observed for $k_2$ of anisotropic dark energy stars using a different model of anisotropy in Ref.~\cite{Pretel:2023nhf}.  
The value of anisotropy parameter $\lambda_{BL}$ has a significant effect on the behaviour of tidal Love number $k_2$ as well. We find that the tidal Love numbers show a significant deviation from isotropy for positive values of $\lambda_{BL}$. Whereas, the deviation observed is comparatively lower for negative values of $\lambda_{BL}$.
The values of the tidal Love number corresponding to the maximum mass ($k_2^{max}$) for $\lambda_{BL}=-2,\,-1,\,0,\,1,$ and $2$ are $0.0147,\,0.0125,\,0.0116,\,0.0098,$ and $0.0079$ respectively. 
Next, we plot the $f$-mode frequencies as a function of tidal deformability $\Lambda$ in Fig.~\ref{fig:L-f}. 
The values of tidal deformability corresponding to the maximum mass ($\Lambda_{max}$) for $\lambda_{BL}=-2,\,-1,\,0,\,1,$ and $2$ are $5.96,\,3.95,\,2.69,\,1.69,$ and $0.97$ respectively. The value of $\Lambda_{max}$ is found to decrease with an increase in $\lambda_{BL}$. 
The values of tidal deformability for $1.4M_\odot$ ($\Lambda_{1.4}$) for $\lambda_{BL}=-2,\,-1,\,0,\,1,$ and $2$ are $49,\,53,\,59,\,60,$ and $76$ respectively.
We find that values of $\Lambda_{1.4}$ are in good agreement with the limit given by GW 170817 ($\Lambda_{1.4}\leq 800$~\cite{LIGOScientific:2017vwq}). However, the values of $f$-mode frequency obtained in our analysis are higher and does not fall within the limit provided in Ref. \cite{Wen:2019ouw}. We hope that future detection of binary NS mergers may provide further constraints on $f$-mode frequencies.

\section{SUMMARY AND CONCLUSIONS}
\label{sec:summ}

We have studied the prominent non-radial $f$-mode oscillations for the isotropic and anisotropic dark energy star configurations. 
By using the modified Chaplygin fluid prescription for the constituent dark energy matter, we solved the respective coupled differential equations in presence of anisotropy for stellar structure, $f$-mode and tidal deformability. Anisotropic pressure of the dark energy fluid is introduced with the Bowers-Liang model, where the free parameter $\lambda_{BL}$ determines the strength of it. 
We find that the internal stellar structure is highly sensitive to the strength of anisotropy and the maximum mass and radius of the star are found to vary significantly with $\lambda_{BL}$.
As $\lambda_{BL}$ increases from $-2$ to $+2$, the maximum mass and corresponding radius of the star are found to increase. Further, we compared the stellar profiles obtained with observational data from GW events and milli-second pulsars and good agreement was found; especially for configurations with $\lambda_{BL} < 0$. Also, the mass-radius profiles of the dark energy star show a different trend from that of the neutron star; while, the curves are qualitatively more similar to that of a quark star. 

The oscillation frequency of $l=2$ $f$-mode of dark energy star was found under the Cowling approximation. We found that the $f-$mode frequencies span the range $\sim 2-3.1$ kHz.  The range of $f$-mode spectrum obtained for the isotropic case was found to be smaller compared to that of neutron and quark stars. 
Further, the effect of anisotropy is reflected in the $f$-mode solutions, where the frequencies show a significant deviation from the isotropic case. This deviation is observed to be high for larger absolute value of the parameter $\lambda_{BL}$, which is expected, since the anisotropic pressure variation is found to be appreciably high for large $|\lambda_{BL}|$ values. Further, the value of $f$ corresponding to the maximum mass of the star, $f_{max}$ is found to decrease with increase in the value of $\lambda_{BL}$. Also, the $f$-mode frequencies were obtained as a function of compactness and tidal deformability. It was found that dark energy stars become ultra compact objects for positive values of $\lambda_{BL}$. It was noted that the tidal deformability corresponding to the maximum mass $\Lambda_{max}$ decreases when $\lambda_{BL}$ is increased. Further, the tidal deformability of anisotropic dark energy stars was found to be in good agreement with the limits imposed by GW 170814 event.

Interestingly, we found that the $f$-mode frequency spectrum of the isotropic and anisotropic dark energy stars is indicatively different from that obtained for neutron stars (NS) and quark stars (QS). 
Further, unlike the case of NS and QS, the $f$-mode spectra of dark energy stars begin within a higher frequency range at $M \gtrsim 1.00\,M_\odot$. We believe that this distinct behaviour of $f$-modes in dark energy stars may give directions in its  identification.

\section*{Acknowledgements}
L. J. N. acknowledges the Department of Science and Technology, Govt. of India for the INSPIRE Fellowship.


\bibliography{DE-fmode}	
\end{document}